\documentstyle[12pt]{article}

\def\DD {{\!}_{DD}}
\def\ap#1#2#3   {{\rm Ann. Phys. (NY)}       {\bf#1}, #2 (#3)}
\def\apj#1#2#3  {{\rm Astrophys. J.}         {\bf#1}, #2 (#3)}
\def\apjl#1#2#3 {{\rm Astrophys. J. Lett.}   {\bf#1}, #2 (#3)}
\def\app#1#2#3  {{\rm Acta. Phys. Pol.}      {\bf#1}, #2 (#3)}
\def\cpc#1#2#3  {{\rm Computer Phys. Comm.}  {\bf#1}, #2 (#3)}
\def\err#1#2#3  {{\it Erratum}               {\bf#1}, #2 (#3)}
\def\ib#1#2#3   {{\it ibid.}                 {\bf#1}, #2 (#3)}
\def\jcp#1#2#3  {{\rm J. Comp. Phys.}        {\bf#1}, #2 (#3)}
\def\jmp#1#2#3  {{\rm J. Math. Phys.}        {\bf#1}, #2 (#3)}
\def\ijmp#1#2#3 {{\rm Int. J. Mod. Phys.}    {\bf#1}, #2 (#3)}
\def\jpg#1#2#3  {{\rm J. Phys. G}            {\bf#1}, #2 (#3)}
\def\mpl#1#2#3  {{\rm Mod. Phys. Lett.}      {\bf#1}, #2 (#3)}
\def\nat#1#2#3  {{\rm Nature (London)}       {\bf#1}, #2 (#3)}
\def\ncim#1#2#3 {{\rm Nuovo Cimento}         {\bf#1}, #2 (#3)}
\def\nim#1#2#3  {{\rm Nucl. Instr. Meth.}    {\bf#1}, #2 (#3)}
\def\np#1#2#3   {{\rm Nucl. Phys.}           {\bf#1}, #2 (#3)}
\def\npb#1#2#3  {{\rm Nucl. Phys. B}         {\bf#1}, #2 (#3)}
\def\pan#1#2#3  {{\rm Phys. At. Nuclei}      {\bf#1}, #2 (#3)}
\def\pl#1#2#3   {{\rm Phys. Lett.}           {\bf#1}, #2 (#3)}
\def\plb#1#2#3  {{\rm Phys. Lett. B}         {\bf#1}, #2 (#3)}
\def\prep#1#2#3 {{\rm Phys. Rep.}            {\bf#1}, #2 (#3)}
\def\prev#1#2#3 {{\rm Phys. Rev.}            {\bf#1}, #2 (#3)}
\def\prd#1#2#3  {{\rm Phys. Rev. D}          {\bf#1}, #2 (#3)}
\def\prl#1#2#3  {{\rm Phys. Rev. Lett.}      {\bf#1}, #2 (#3)}
\def\prs#1#2#3  {{\rm Proc. Roy. Soc.}       {\bf#1}, #2 (#3)}
\def\ptp#1#2#3  {{\rm Prog. Theor. Phys.}    {\bf#1}, #2 (#3)}
\def\ps#1#2#3   {{\rm Physica Scripta}       {\bf#1}, #2 (#3)}
\def\rmp#1#2#3  {{\rm Rev. Mod. Phys.}       {\bf#1}, #2 (#3)}
\def\rpp#1#2#3  {{\rm Rep. Prog. Phys.}      {\bf#1}, #2 (#3)}
\def\sjnp#1#2#3 {{\rm Sov. J. Nucl. Phys.}   {\bf#1}, #2 (#3)}
\def\spj#1#2#3  {{\rm Sov. Phys. JETP}       {\bf#1}, #2 (#3)}
\def\spjl#1#2#3 {{\rm Sov. JETP Lett.}       {\bf#1}, #2 (#3)}
\def\spu#1#2#3  {{\rm Sov. Phys.-Usp.}       {\bf#1}, #2 (#3)}
\def\zp#1#2#3   {{\rm Zeit. Phys.}           {\bf#1}, #2 (#3)}
\def\zpc#1#2#3  {{\rm Z. Phys. C}            {\bf#1}, #2 (#3)}
\begin{document}
  \begin{center}
  \begin{Large} \begin{bf}
    HIPPOPO \\ a Monte Carlo generator for particle production
    \\in the Quark Gluon String Model
  \end{bf} \end{Large} \end{center}
  \begin{center}
  S.P.~Baranov\\
 {\sl P.N.Lebedev~Institute~of~Physics,~Moscow,~Russia}\\
 {\sl E-mail: baranov@sci.lebedev.ru}\\~\\
  \end{center}
{\Large{\bf 1~~Introduction}}\\~\\
HIPPOPO (Hadron Induced Particle Production Off Pomeron Outpour)
is a program to simulate the production of hadrons via the
multipomeron exchange mechanism in a high energy hadron collision.
The underlying algorithm is based upon the Quark Gluon String Model
developed over the years by many authors \cite{Cap70}-\cite{Ara95}.

For the initial colliding beams the User can choose among the following
species: nucleons, atomic nuclei, $\pi$, $K$, $\phi$ and $J/\psi$ mesons,
$\Lambda$, $\Sigma$ and $\Xi$ hyperons (of any explicitly stated charge),
and their antiparticles. Also, the program allows photon beams. In the 
latter case, the photon is treated as a superposition of $\rho$, $\phi$
and $J/\psi$ mesons according to the Vector Dominance Model. 

The produced hadron may be a $\pi$, $\eta$, $\rho$, $\omega$, $K$, $\phi$, 
$D$, $D_s$ or $J/\psi$ meson, nucleon $N$, $\Lambda$, $\Sigma$, $\Xi$ or 
$\Omega$ hyperon, $\Lambda_c$ or $\Xi_c$ baryon, an antiparticle or
an excited state of any mentioned hadron.

The program yields the fully differential inclusive cross section. However,
it is not a full event generator, because it only focuses on the inclusive
production properties of the user-defined hadron, while the accompanying
particles are not taken into account.
\\~\\
{\Large{\bf 2~~Physics input}}\\
\par
An interaction between the colliding hadrons is assumed to proceed by means
of the multipomeron exchange mechanism. The particular contribution to the 
cross section due to the exchange of $n$ Pomerons reads \cite{Ter73}:
\begin{equation}
\sigma_n(\xi)=\frac{\sigma_P}{nz}\,\Bigl( 1-e^{{\large -z}}\,
\sum_{k=0}^{n-1}\,\frac{z^k}{k!} \Bigr) ,\qquad n\ge 1 \label{sigmaN}
\end{equation}
with
$$
\xi=\ln{s},\qquad \sigma_P=8\pi\gamma_P\,e^{{\large \xi\Delta}},\qquad
z=2C\gamma_P\,[R^2+\alpha'_P\xi]^{-1}\,s^{{\large \Delta}},
$$
and $C$, $\gamma_P$, $R^2$, $\alpha'_P$ and $\Delta$ being the
model parameters (see section 3.3 for their numerical values).
Also, there exist contributions from the elastic and diffraction dissociation 
processes, $\sigma_{el}$ and $\sigma_{\DD}$, which correspond to the
absense of Pomerons ($n=0$):
\begin{equation}
\sigma_0(\xi)=\sigma_P \bigl[ f(\frac{z}{2})-f(z) \bigr] \quad \label{sigma0}
\mbox{with}\quad f(z)=\sum_{\nu=1}^{\infty}\,\frac{(-z)^{\nu-1}}{\nu\;\nu !},
\end{equation}
\begin{equation}
\sigma_{el}=(1/C)\;\sigma_0(\xi),\qquad
\sigma_{\DD}=(1-1/C)\;\sigma_0(\xi).
\end{equation}
Summing up, the total inelastic cross section is:
\begin{equation}
\sigma_{tot}=\sigma_{\DD} + \sum_{n=1}^{\infty}\,\sigma_n \label{sigmaT}.
\end{equation}

According to the Quark Gluon String Model, each Pomeron is treated as a pair
of colour strings attached to the partons in the colliding hadrons. The
fragmentation of colour strings results in the production of particles.
The inclusive differential cross section for a hadron of type $h$ reads:
\begin{equation}
\frac{d\sigma^h}{d^2p_T\,dy}=\sigma_{\DD}\varphi_{\DD}^h +
 \sum_{n=1}^{\infty}\,\sigma_n\varphi_n^h \label{sigmaH},
\end{equation}
where the particle production from Pomeron strings is described by the
string fragmentation functions $\varphi_n$:
\begin{eqnarray}
\varphi_n^h&=&a^h\,\bigl[ F_{val}^h(x_{+},n)F_{\overline{val}}^h(x_{-},n) \label{phi}
                      +F_{\overline{val}}^h(x_{+},n)F_{val}^h(x_{-},n) \nonumber \\
      &&        +(n-1)[F_{sea}^h(x_{+},n)F_{\overline{sea}}^h(x_{-},n)
                      +F_{\overline{sea}}^h(x_{+},n)F_{sea}^h(x_{-},n)]\bigr],
\end{eqnarray}
with $x_{\pm}=\frac{1}{2}[(x_{\perp}^2+x_F^2)^{1/2} \pm x_F],\quad
      x_{\perp}=2m^h_{\perp}/\sqrt{s},$ \quad so that \quad
     $x_{+}x_{-}=(m^h_{\perp})^2/s$ \quad and \quad $x_{+}-x_{-}=x_F.$\\
Each term in (\ref{phi}) corresponds to an independent string stretched
between oppositely coloured partons from different beams.
The contribution from each string is given by a product of two independent
factors, $F_i(x_{+})$ and $F_{\bar i}(x_{-})$, related to the endpoint
partons (moving in the positive and negative directions, respectively).
In turn, the factors $F(x_{\pm})$ are given by the convolution
of universal functions:
\begin{equation}  \label{F}
F^h(x_{\pm},n)=\sum_i
\int_{x_{\pm}}^1 f_i(x',n)\,G_i^h(x_{\pm}/x',\,p_T)\,T(x_F/x',\,p_T,\,n)\,dx',
\end{equation}
where $f_i(x,n)$ stands for the probability to find a parton of type $i$
carrying the momentum fraction $x$ in the beam hadron,
and $G_i^h(x,p_{\perp})$ stands for the probability that the fragmentation of
a type $i$ parton yields a hadron $h$ with the longitudinal momentum
fraction $x$ and transverse momentum $p_{\perp}$.
The parton distribution functions $f_i(x,n)$ are specific for a given beam
particle, and the fragmentation functions $G_i^h(x,p_{\perp})$ are specific
for a given outstate hadron.

The diffraction dissociation term was taken in the form \cite{Pis86}
\begin{equation}  \label{DDbaryon}
\varphi_{\DD}^h=\sqrt{x_{+}}F^h(x_{+},1)+\sqrt{x_{-}}F^h(x_{-},1)
\end{equation}
if the fragmenting and the produced particles are both baryons or both
antibaryons, and
\begin{equation}  \label{DDmeson}
\varphi_{\DD}^h=(3/2)[\sqrt{x_{+}}+\sqrt{x_{-}}]\,F^h(x_{+},1)\,F^h(x_{-},1)
\end{equation}
in all other cases.

The parton distribution functions have the (Reggeon theory inspired) generic form:
\begin{equation}
f(x,n)=x^{\alpha_f}(1-x)^{\beta_f+n-1} \label{f}
\end{equation}
with the exponents $\alpha_f$ and $\beta_f$ determined by the flavour content
of the beam particle. Similarly, the generic form of the parton fragmentation
functions is:
\begin{equation}
G(x,p_T)=x^{\alpha_G}(1-x)^{\beta_G+2\alpha'_Rp_T^2} \label{G}
\end{equation}
with the exponents $\alpha_G$ and $\beta_G$ determined by the flavour content
of the produced hadron. The parameters $\alpha_f$, $\beta_f$, $\alpha_G$ and
$\beta_G$ are expressed in terms of the relevant Regge intercepts and
represent the basic model assumptions.
At the same time, to provide an interpolation between the limiting points
$x\to 0$ and $x\to 1$, some extra factors may be introduced in the functions
$G_i^h(x,p_T)$. The latter ones are only of phenomenological meaning and cannot
be derived from `first principles'. The full list of parton distribution 
and fragmentation functions is too long to be presented in this paper, but 
these functions are easily readable from the FORTRAN code.

Finally, the weight function $T(z,p_T,n)$ in (\ref{F}) is present for the
hadron transverse momentum distribution. According to \cite{Ves85}, it is
parametrized in the form:
\begin{equation}
T(z,p_T,n)=[b_z^2/2\pi(1+b_zm^h)]\,\exp[-b_z(m_{\perp}^h-m^h)] \label{T}
\end{equation}
with $b_z=\gamma^h/[1+(2-1/n)\rho z^2]$, and $\gamma^h$ and $\rho$ being
still new phenomenological parameters.\\~\\~\\
{\Large{\bf 3~~Program components}}\\~\\
{\large{\bf 3.1~General structure and external references}}\\
\par The program structure is divided into several pieces. These are
the user job cards and the initiating routines 
 (collected in the file {\tt run.f});
the table of parton distribution functions (file {\tt parton.f});
the table of fragmentation functions 
 (files {\tt fragm1.f}, {\tt fragm2.f}, {\tt fragm3.f});
the particle production algorithm as described in Sect. 2
 (file {\tt model.f});
the Monte Carlo integration routine VEGAS \cite{VEGAS} with its intrinsic
random number generator (file {\tt vegas.f}).

When evaluating the convolution in (\ref{F}) the program refers to the standard
function {\tt DGAUSS} from CERN library. The normalizing factors for parton
distribution functions are calculated with the use of the {\tt DGAMMA} function,
also from CERN library.

For histogramming purposes, the program uses the standard PAW package.
%
%
\\~\\
{\large{\bf 3.2~Subroutines and Functions}}\\~\\
{\tt  SUBROUTINE CKNAME(i1,i2,ih)}\\
\hspace*{5mm}\parbox{15.5cm}{Checks the consistency of the user given names
      for the beam and the produced particles.}\\
{\tt  SUBROUTINE PARAM}\\
\hspace*{5mm}\parbox{15.5cm}{Introduces the model parameters.}\\
{\tt  SUBROUTINE HYBRID}\\
\hspace*{5mm}\parbox{15.5cm}{Assignes model dependent intercepts
      to hybrid and excited states.}\\
{\tt  SUBROUTINE MIX}\\
\hspace*{5mm}\parbox{15.5cm}{An algorithm to include composite beams (Atomic
      nuclei, for example) and hadrons of indefinite quark content (short and
      long living Kaons, $\eta'$ mesons, etc.).}\\~\\
{\tt  SUBROUTINE CXNPOM(sqs,nmax)}\\
\hspace*{5mm}\parbox{15.5cm}{Calculates the multipomeron cross sections, equ. (\ref{sigmaN}).}\\
{\tt  FUNCTION F0(z)}\\
\hspace*{5mm}\parbox{15.5cm}{Returns the $f(z)$ value, equ. (\ref{sigma0}).}\\~\\
%
{\tt  FUNCTION FXN(xveg,wgt)}\\
\hspace*{5mm}\parbox{15.5cm}{The integrand expression for VEGAS \cite{VEGAS},
      with {\tt xveg} being the array of phase space variables and {\tt wgt}
      the weight factor due to the particular phase space binning (supplied
      by VEGAS).}\\~\\
{\tt  FUNCTION F1dif(x), F2dif(x)}\\
\hspace*{5mm}\parbox{15.5cm}{1st and 2nd beam diffractive dissociation functions.}\\
{\tt  FUNCTION F1val(x), F2val(x)}\\
\hspace*{5mm}\parbox{15.5cm}{1st and 2nd beam valent colour endpoint functions.}\\
{\tt  FUNCTION F1bar(x), F2bar(x)}\\
\hspace*{5mm}\parbox{15.5cm}{1st and 2nd beam valent anticolour endpoint functions.}\\
{\tt  FUNCTION F1v(x), F2v(x)}\\
\hspace*{5mm}\parbox{15.5cm}{1st and 2nd beam sea colour endpoint functions.}\\
{\tt  FUNCTION F1b(x), F2b(x)}\\
\hspace*{5mm}\parbox{15.5cm}{1st and 2nd beam sea anticolour endpoint functions.}\\
{\tt  FUNCTION T(zF,n,hmass,pt2)}\\
\hspace*{5mm}\parbox{15.5cm}{Transverse momentum distribution function.}\\~\\
{\tt  SUBROUTINE NORM(nmax)}\\
\hspace*{5mm}\parbox{15.5cm}{Normalizes all parton distribution functions.}\\
{\tt  FUNCTION Bfun(ex,e1)}\\
\hspace*{5mm}\parbox{15.5cm}{Euler's Beta function.}\\~\\
{\tt  FUNCTION}\\
{\tt  FvalU(x,n,beam), FvalD(x,n,beam), FvalS(x,n,beam), FvalC(x,n,beam)}\\
\hspace*{5mm}\parbox{15.5cm}{Valent $u$, $d$, $s$, $c$ quark distribution functions.}\\
{\tt  FUNCTION}\\
{\tt  FbarU(x,n,beam), FbarD(x,n,beam), FbarS(x,n,beam), FbarC(x,n,beam)}\\
\hspace*{5mm}\parbox{15.5cm}{Valent $\overline{u\strut}$, $\overline{d\strut}$,
      $\overline{s\strut}$, $\overline{c\strut}$ antiquark distribution functions.}\\
{\tt  FUNCTION}\\
{\tt  FseaU(x,n,beam), FseaD(x,n,beam), FseaS(x,n,beam), FseaC(x,n,beam)}\\
\hspace*{5mm}\parbox{15.5cm}{Sea $u$, $d$, $s$, $c$ quark distribution functions,
                              apply to antiquarks also.}\\
{\tt  FUNCTION}\\
{\tt  FvalUU(x,n,beam), FvalUD(x,n,beam), FvalDD(x,n,beam)}\\
\hspace*{5mm}\parbox{15.5cm}{Valent $uu$, $ud$, $dd$ diquark distribution functions.}\\
{\tt  FUNCTION}\\
{\tt  FvalUS(x,n,beam), FvalDS(x,n,beam), FvalSS(x,n,beam)}\\
\hspace*{5mm}\parbox{15.5cm}{Valent $us$, $ds$, $ss$ diquark distribution functions.}\\
{\tt  FUNCTION}\\
{\tt  FbarUU(x,n,beam), FbarUD(x,n,beam), FbarDD(x,n,beam)}\\
\hspace*{5mm}\parbox{15.5cm}{Valent $\overline{uu\strut}$, $\overline{ud\strut}$,
      $\overline{dd\strut}$ antidiquark distribution functions.}\\
{\tt  FUNCTION}\\
{\tt  FbarUS(x,n,beam), FbarDS(x,n,beam), FbarSS(x,n,beam)}\\
\hspace*{5mm}\parbox{15.5cm}{Valent $\overline{us\strut}$, $\overline{ds\strut}$,
      $\overline{ss\strut}$ antidiquark distribution functions. In the above functions:}\\
\hspace*{5mm}\parbox{1.2cm}{{\tt x   }}\parbox{13cm}{is the momentum fraction,}\\
\hspace*{5mm}\parbox{1.2cm}{{\tt n   }}\parbox{13cm}{is the number of exchanged Pomerons,}\\
\hspace*{5mm}\parbox{1.2cm}{{\tt beam}}\parbox{13cm}{is the beam hadron the distribution functions refer to.}\\~\\
{\tt  FUNCTION}\\
{\tt  DvalU(x,hadr,pt2), DvalD(x,hadr,pt2), DvalS(x,hadr,pt2), DvalC(x,hadr,pt2)}\\
\hspace*{5mm}\parbox{15.5cm}{The $u$, $d$, $s$, $c$ quark diffraction dissociation functions.}\\
{\tt  FUNCTION}\\
{\tt  DbarU(x,hadr,pt2), DbarD(x,hadr,pt2), DbarS(x,hadr,pt2), DbarC(x,hadr,pt2)}\\
\hspace*{5mm}\parbox{15.5cm}{The $\overline{u\strut}$, $\overline{d\strut}$,
      $\overline{s\strut}$, $\overline{c\strut}$ antiquark diffraction dissociation functions.}\\
{\tt  FUNCTION}\\
{\tt  DvalUU(x,hadr,pt2), DvalUD(x,hadr,pt2), DvalDD(x,hadr,pt2)}\\
\hspace*{5mm}\parbox{15.5cm}{The $uu$, $ud$, $dd$ diquark diffraction dissociation functions.}\\
{\tt  FUNCTION}\\
{\tt  DvalUS(x,hadr,pt2), DvalDS(x,hadr,pt2), DvalSS(x,hadr,pt2)}\\
\hspace*{5mm}\parbox{15.5cm}{The $us$, $ds$, $ss$ diquark diffraction dissociation functions.}\\
{\tt  FUNCTION}\\
{\tt  DbarUU(x,hadr,pt2), DbarUD(x,hadr,pt2), DbarDD(x,hadr,pt2)}\\
\hspace*{5mm}\parbox{15.5cm}{The $\overline{uu\strut}$, $\overline{ud\strut}$,
      $\overline{dd\strut}$ antidiquark diffraction dissociation functions.}\\
{\tt  FUNCTION}\\
{\tt  DbarUS(x,hadr,pt2), DbarDS(x,hadr,pt2), DbarSS(x,hadr,pt2)}\\
\hspace*{5mm}\parbox{15.5cm}{The $\overline{us\strut}$, $\overline{ds\strut}$,
      $\overline{ss\strut}$ antidiquark diffraction dissociation functions.}\\~\\
{\tt  FUNCTION}\\
{\tt  GvalU(x,hadr,pt2), GvalD(x,hadr,pt2), GvalS(x,hadr,pt2), GvalC(x,hadr,pt2)}\\
\hspace*{5mm}\parbox{15.5cm}{The $u$, $d$, $s$, $c$ quark fragmentation functions.}\\
{\tt  FUNCTION}\\
{\tt  GbarU(x,hadr,pt2), GbarD(x,hadr,pt2), GbarS(x,hadr,pt2), GbarC(x,hadr,pt2)}\\
\hspace*{5mm}\parbox{15.5cm}{The $\overline{u\strut}$, $\overline{d\strut}$,
      $\overline{s\strut}$, $\overline{c\strut}$ antiquark fragmentation functions.}\\
{\tt  FUNCTION}\\
{\tt  GvalUU(x,hadr,pt2), GvalUD(x,hadr,pt2), GvalDD(x,hadr,pt2)}\\
\hspace*{5mm}\parbox{15.5cm}{The $uu$, $ud$, $dd$ diquark fragmentation functions.}\\
{\tt  FUNCTION}\\
{\tt  GvalUS(x,hadr,pt2), GvalDS(x,hadr,pt2), GvalSS(x,hadr,pt2)}\\
\hspace*{5mm}\parbox{15.5cm}{The $us$, $ds$, $ss$ diquark fragmentation functions.}\\
{\tt  FUNCTION}\\
{\tt  GbarUU(x,hadr,pt2), GbarUD(x,hadr,pt2), GbarDD(x,hadr,pt2)}\\
\hspace*{5mm}\parbox{15.5cm}{The $\overline{uu\strut}$, $\overline{ud\strut}$,
      $\overline{dd\strut}$ antidiquark fragmentation functions.}\\
{\tt  FUNCTION}\\
{\tt  GbarUS(x,hadr,pt2), GbarDS(x,hadr,pt2), GbarSS(x,hadr,pt2)}\\
\hspace*{5mm}\parbox{15.5cm}{The $\overline{us\strut}$, $\overline{ds\strut}$,
      $\overline{ss\strut}$ antidiquark fragmentation functions. In the above functions:}\\
\hspace*{5mm}\parbox{1.2cm}{{\tt x   }}\parbox{13cm}{is the momentum fraction,}\\
\hspace*{5mm}\parbox{1.2cm}{{\tt hadr}}\parbox{13cm}{is the hadron to appear from fragmentation,}\\
\hspace*{5mm}\parbox{1.2cm}{{\tt pt2 }}\parbox{13cm}{is its transverse momentum squared.}\\~\\
{\tt  SUBROUTINE DDMODE}\\
\hspace*{5mm}\parbox{15.5cm}{Establishes the incoming and outcoming particle types as 'meson',
      'baryon' or 'antibaryon', that is important for diffraction dissociation contributions.}\\~\\
{\tt  SUBROUTINE ABSRAT}\\
\hspace*{5mm}\parbox{15.5cm}{Establishes the absolute production rate.}\\
\newpage \noindent
{\tt  SUBROUTINE VEGAS(FXN,AVGI,SD,CHI2A)}\\
\hspace*{5mm}\parbox{15.5cm}{Perform multidimensional Monte Carlo
      integration \cite{VEGAS}.}\\
{\tt  DOUBLE PRECISION FUNCTION RANDOM(SEED)}\\
\hspace*{5mm}\parbox{15.5cm}{Random number generator \cite{RANDOM}.}\\~\\
{\tt  SUBROUTINE WRIOUT(hwgt)}\\
\hspace*{5mm}\parbox{15.5cm}{A user supplied routine to manage output
  information. {\tt hwgt} is the correctly normalized weight that
  reproduces the physical cross section (\ref{sigmaH}) when is summed
  over all the generated events.}\\~\\~\\
{\large{\bf 3.3~Common blocks}}\\~\\
{\tt  COMMON/INPUT/beam1,beam2,hadron} (All are {\tt CHARACTER*6} names)\\
\hspace*{5mm}\parbox{2.2cm}{{\tt beam1 }}\parbox{13cm}{1st beam particle type (defined by the User).}\\
\hspace*{5mm}\parbox{2.2cm}{{\tt beam2 }}\parbox{13cm}{2nd beam particle type (defined by the User).}\\
\hspace*{5mm}\parbox{2.2cm}{{\tt hadron}}\parbox{13cm}{The produced hadron type (defined by the User).}\\
{\tt  COMMON/BEAMS/beam1,beam2,hadron} (All are {\tt CHARACTER*6} names)\\
\hspace*{5mm}\parbox{2.2cm}{{\tt beam1 }}\parbox{13cm}{1st beam particle type (for internall use).}\\
\hspace*{5mm}\parbox{2.2cm}{{\tt beam2 }}\parbox{13cm}{2nd beam particle type (for internall use).}\\
\hspace*{5mm}\parbox{2.2cm}{{\tt hadron}}\parbox{13cm}{The produced hadron type (for internall use).}\\
{\tt  COMMON/ATOMN/Atom1(2),Atom2(2)}\\
\hspace*{5mm}\parbox{2.2cm}{{\tt Atom1(1)}}\parbox{13cm}{Atomic number of the 1st beam (if nucleus).}\\
\hspace*{5mm}\parbox{2.2cm}{{\tt Atom1(2)}}\parbox{13cm}{Atomic weight of the 1st beam (if nucleus).}\\
\hspace*{5mm}\parbox{2.2cm}{{\tt Atom2(1)}}\parbox{13cm}{Atomic number of the 2nd beam (if nucleus).}\\
\hspace*{5mm}\parbox{2.2cm}{{\tt Atom2(2)}}\parbox{13cm}{Atomic weight of the 2nd beam (if nucleus).}\\
{\tt  COMMON/EXCIT/Level}\\
\hspace*{5mm}\parbox{15.5cm}{The outgoing hadron excitation level, {\tt Level=0} is the ground state.}\\
{\tt  COMMON/VDMES/M1(3),M2(3)}\\
\hspace*{5mm}\parbox{15.5cm}{The keys that switch on and off the vector meson component in a photon.
                        {\tt M1(i)} and {\tt M2(i)} refer to the 1st and the 2nd beams, respectively,
                        and {\tt i=1,2,3} stand for $\rho$, $\phi$ and $J/\psi$ meson contributions.}\\~\\
{\tt  COMMON/KINEM/sqs,x1,x2,xF,y,hmass,pt2}\\
\hspace*{5mm}\parbox{2.2cm}{{\tt sqs  }}\parbox{13cm}{Total c.m.s. energy, $\sqrt{s}$ [GeV].}\\
\hspace*{5mm}\parbox{2.2cm}{{\tt x1   }}\parbox{13cm}{Parton momentum fraction in the 1st beam, $x_{+}$}\\
\hspace*{5mm}\parbox{2.2cm}{{\tt x2   }}\parbox{13cm}{Parton momentum fraction in the 2nd beam, $x_{-}$}\\
\hspace*{5mm}\parbox{2.2cm}{{\tt xF   }}\parbox{13cm}{Feynman variable of the produced hadron, $x_F$.}\\
\hspace*{5mm}\parbox{2.2cm}{{\tt y    }}\parbox{13cm}{Rapidity of the produced hadron, $y$.}\\
\hspace*{5mm}\parbox{2.2cm}{{\tt hmass}}\parbox{13cm}{Mass of the produced hadron, $m_H$ [GeV].}\\
\hspace*{5mm}\parbox{2.2cm}{{\tt pt2  }}\parbox{13cm}{Transverse momentum squared, $p_T^2$ [GeV$^2$].}\\
{\tt  COMMON/STRNG/n,nmax}\\
\hspace*{5mm}\parbox{15.5cm}{The number of Pomerons exchanged and the maximal number of Pomerons}\\
{\tt  COMMON/POMER/Coeff,Gammp,Rpom2,AlPom,Delta}\\
\hspace*{5mm}\parbox{3.2cm}{{\tt Coeff = 1.5 }}\parbox{12cm}{$C=1+\sigma_{\DD}/\sigma_{el}$, equ.(\ref{sigmaN})}\\
\hspace*{5mm}\parbox{3.2cm}{{\tt Gammp = 3.64}}\parbox{12cm}{Pomeron residue parameter, $\gamma_P$ [GeV$^{-2}$]}\\
\hspace*{5mm}\parbox{3.2cm}{{\tt Rpom2 = 3.56}}\parbox{12cm}{Pomeron size parameter, $R^2$ [GeV$^{-2}$]}\\
\hspace*{5mm}\parbox{3.2cm}{{\tt AlPom = 0.25}}\parbox{12cm}{Pomeron trajectory slope, $\alpha'_P$ [GeV$^{-2}$]}\\
\hspace*{5mm}\parbox{3.2cm}{{\tt Delta = 0.07}}\parbox{12cm}{Pomeron intercept criticality, $\Delta$}\\
\hspace*{5mm}\parbox{15.5cm}{The above parameters are fixed in the model
                             and are not recommended to be altered.}\\~\\
{\tt  COMMON/CXPOM/CXN(0:NP)}\\
\hspace*{5mm}\parbox{15.5cm}{The multipomeron exchange cross sections $\sigma_{\DD}$, $\sigma_n$.}\\
{\tt  COMMON/REGGE/alp,aR,aP,aJ,aN,aD,aV,aX,aO,aE}\\
\hspace*{5mm}\parbox{15.5cm}{The intercept parameters for Reggeon, $\rho$,
  $\varphi$, $J/\psi$, $N$, $\Delta$, $\Lambda$, $\Xi$, $\Omega$ trajectories.
  Please note that the parameters {\tt aN, aD, aV, aX, aO} must not be
  literally copied from Particle Data. They do not have actual meaning of
  baryon intercepts, but only enter the model in artificial combinations that
  correspond to fictituous trajectories of multiquark hybrid states.
  {\tt aE} is the effective intercept shift for excited states (a model).}\\
{\tt  COMMON/TRNSV/gamh,gamr}\\
\hspace*{5mm}\parbox{15.5cm}{Transverse momentum distribution parameters
  $\gamma^h$ and $\rho$, see equ. (\ref{T}).}\\
{\tt  COMMON/STRAN/Snucl,Cnucl}\\
\hspace*{5mm}\parbox{15.5cm}{Strange and Charm sea suppression parameters.}\\
{\tt  COMMON/PIPHI/Cpi,Cph,Cps}\\
\hspace*{5mm}\parbox{15.5cm}{Fragmentation parameters for $\pi$ and $\phi$ mesons.
        Note that only the associated production of the type $\phi K\bar K$
        is considered for $\phi$ mesons. The OZI-violating nonplanar diagrams
        are not included at present.}\\
{\tt  COMMON/PNLAM/Cn1,Cn2,Cl1,Cl2}\\
\hspace*{5mm}\parbox{15.5cm}{Fragmentation parameters for $p$, $n$ and $\Lambda$.}\\
{\tt  COMMON/KAONS/Ck,Cks,Ck1,Ck2}\\
\hspace*{5mm}\parbox{15.5cm}{Fragmentation parameters for Kaons.}\\
{\tt  COMMON/HYPER/Cs1,Cs2,Cx1,Cx2,Co1,Co2}\\
\hspace*{5mm}\parbox{15.5cm}{Fragmentation parameters for $\Sigma$, $\Xi$ and $\Omega$ hyperons.}\\
{\tt  COMMON/CHARM/Cd,Cd1,Cd2,Cc1,Cc2,Cf,Cfs,Cfc,Cj,Cjc}\\
\hspace*{5mm}\parbox{15.5cm}{Fragmentation parameters for $D$, $D_s$, $J/\psi$ 
        mesons and $\Lambda_c$ baryon.
        Note that only the associated production of the type $J/\psi D\bar D$
        is considered for $J/\psi$ mesons. The OZI-violating nonplanar diagrams
        are not included at present.}\\
{\tt  COMMON/SUPER/Cxc,Cxc1,Cxc2}\\
\hspace*{5mm}\parbox{15.5cm}{Fragmentation parameters for $\Xi_c$ baryons.}\\
{\tt  COMMON/VDPAR/VDM(3)}\\
\hspace*{5mm}\parbox{15.5cm}{Photon to vector meson coupling constannts, $4\pi\alpha/f_V^2$.}\\
{\tt  COMMON/RATES/rrho,reta,retap,romm,rexcit}\\
{\tt  COMMON/YIELD/Ctotal}\\
\hspace*{5mm}\parbox{15.5cm}{Normalization parameters for hadron production rates.}\\
{\tt  COMMON/INDEX/ex,e1}\\
\hspace*{5mm}\parbox{15.5cm}{The exponents that parameterize parton distributions.}\\
\hspace*{-2mm}\parbox{15cm}{%
{\tt \hspace*{2.2mm}COMMON/CNORM/CvalU(2,NP), CvalD(2,NP), CvalS(2,NP), CvalC(2,NP)}\\
{\tt .\hphantom{COMMON/CNORM},CbarU(2,NP), CbarD(2,NP), CbarS(2,NP), CbarC(2,NP)}\\
{\tt .\hphantom{COMMON/CNORM},CseaU(2,NP), CseaD(2,NP), CseaS(2,NP), CseaC(2,NP)}\\
{\tt .\hphantom{COMMON/CNORM},CvalUU(2,NP),CvalUD(2,NP),CvalDD(2,NP)}\\
{\tt .\hphantom{COMMON/CNORM},CvalUS(2,NP),CvalDS(2,NP),CvalSS(2,NP)}\\
{\tt .\hphantom{COMMON/CNORM},CbarUU(2,NP),CbarUD(2,NP),CbarDD(2,NP)}\\
{\tt .\hphantom{COMMON/CNORM},CbarUS(2,NP),CbarDS(2,NP),CbarSS(2,NP)}}\\
\hspace*{5mm}\parbox{15.5cm}{The normalization factors for parton distributions
 in the 1st and the 2nd beams, for an $n$-Pomeron exchange.}\\~\\
\newpage \noindent
{\tt  COMMON/BVEG1/XL(10),XU(10),ACC}\\
\hspace*{5mm}\parbox{2.2cm}{{\tt XL(10)}}\parbox{13cm}{Array of lower limits for phase space variables}\\
\hspace*{5mm}\parbox{2.2cm}{{\tt XU(10)}}\parbox{13cm}{Array of upper limits for phase space variables}\\
\hspace*{5mm}\parbox{2.2cm}{{\tt ACC   }}\parbox{13cm}{Accuracy parameter}\\
{\tt  COMMON/BVEGG/NDIM,NCALL,ITMX,NPRN}\\
\hspace*{5mm}\parbox{2.2cm}{{\tt NDIM }}\parbox{13cm}{Number of phase space dimensions}\\
\hspace*{5mm}\parbox{2.2cm}{{\tt NCALL}}\parbox{13cm}{Number of random points per iteration}\\
\hspace*{5mm}\parbox{2.2cm}{{\tt ITMX }}\parbox{13cm}{Maximal number of iterations}\\
\hspace*{5mm}\parbox{2.2cm}{{\tt NPRN }}\parbox{13cm}{Print/noprint parameter}\\
{\tt  COMMON/SEED/NUM}\\
\hspace*{5mm}\parbox{2.2cm}{{\tt NUM}} \parbox{13cm}{Random number}\\~\\
%
{\large{\bf 3.4~Job cards}}\\
\par To run the program, the User
has first to initiate the random number generator by the card {\tt NUM=1},
to set the total c.m.s. energy and to specify the colliding beams and the
hadron to be produced.

Also, the User has to define the lower {\tt XL(i)} and the upper {\tt XU(i)}
limits for the independent variables that parametrize the phase space, i.e.
$\ln{p_T^2}$ and $y$. (Note the use of logarithm of the transverse momentum.)
The lower limit of the transverse momentum is an arbitrary small number (it
should be only nonzero, to avoid formal arithmetical conflicts). The upper
limit should be chosen in a reasonable agreement with the total c.m.s. energy
{\tt sqs}.

The wanted histograms have also to be booked in the {\tt MAIN} program.

It is recommended to perform calculations in two steps. A short preliminary
run optimizes the VEGAS grid to the integrand function shape:\\
\tt
\parbox{14mm}{~}\parbox{4cm}{NCALL = 1000   }\parbox{9cm}{! number of points per iteration}\\
\parbox{14mm}{~}\parbox{4cm}{ITMX  = 5      }\parbox{9cm}{! number of iterations          }\\
\parbox{14mm}{~}\parbox{4cm}{NPRN  = 0      }\parbox{9cm}{! do not fill histograms        }\\
\parbox{14mm}{~}\parbox{9cm}{CALL VEGAS(FXN,AVGI,SD,CHI2A)}\\
\rm
After that one can start a long run to accumulate large statistics:\\
\tt
\parbox{14mm}{~}\parbox{4cm}{NCALL = 200000 }\parbox{9cm}{! number of points per iteration}\\
\parbox{14mm}{~}\parbox{4cm}{ITMX  = 1      }\parbox{9cm}{! number of iterations          }\\
\parbox{14mm}{~}\parbox{4cm}{NPRN  = 1      }\parbox{9cm}{! do fill histograms            }\\
\parbox{14mm}{~}\parbox{9cm}{CALL VEGAS1(FXN,AVGI,SD,CHI2A)}\\
\rm
The quantity that is to be plotted in histograms for the physical cross
section is given by {\tt hwgt} (see {\tt SUBROUTINE WRIOUT}).\\~\\
%
{\large{\bf Acknowledgements}}\\
\par The creation of the HIPPOPO code was supported by Deutsche
Forschungsgemeinschaft, contract number DFG 436 RUS 113/465/0-2(R),
and Russian Foundation for Basic Research, grant number RFFI 00-02-04018.
The code is public and is available from the author on request.
%

\end{document}